\documentclass[prb,aps,nobalancelastpage,amssymb,twocolumn]{revtex4}

\usepackage{graphicx}

\begin{document}

\title{Optical response of high-$T_c$ cuprates:\\
possible role of scattering rate saturation and in-plane anisotropy}

\author{N. E. Hussey, J. C. Alexander \& R. A. Cooper }

\affiliation{H. H. Wills Physics Laboratory, University of Bristol, Tyndall Avenue, BS8 1TL, U.K.}

\date{\today}

\begin{abstract}
We present a generalized Drude analysis of the in-plane optical conductivity $\sigma_{ab}$($T$,$\omega$) in cuprates
taking into account the effects of in-plane anisotropy. A simple ansatz for the scattering rate $\Gamma$($T$,$\omega$),
that includes anisotropy, a quadratic frequency dependence and saturation at the Mott-Ioffe-Regel limit, is able to
reproduce recent normal state data on an optimally doped cuprate over a wide frequency range. We highlight the
potential importance of including anisotropy in the full expression for $\sigma_{ab}$($T$,$\omega$) and challenge
previous determinations of $\Gamma$($\omega$) in which anisotropy was neglected and $\Gamma$($\omega$) was indicated to
be strictly linear in frequency over a wide frequency range. Possible implications of our findings for understanding
thermodynamic properties and self-energy effects in high-$T_c$ cuprates will also be discussed.
\end{abstract}

\maketitle

\section{Introduction}
\label{intro}

The normal state in-plane charge dynamics of cuprate superconductors, both in- and out-of-plane, are still poorly
understood despite two decades of intensive research.\cite{HusseyReview} While the experimental situation is now well
established, its theoretical interpretation remains controversial, largely due to the high transition temperatures
themselves restricting the temperature range over which individual models can be critically examined. In this regard,
measurements of the in-plane optical conductivity $\sigma_{ab}$($T$,$\omega$) play a central role. The ability to
resolve small spectral weight differences between the normal and superconducting state (the so-called
Ferrell-Glover-Tinkham sum rule) is testimony to the improvement in quality of optical conductivity data (and its
analysis) in recent years. \cite{Molegraaf, Santander-Syro, Boris} This has also led to an extension of the energy
scale (up to the bare bandwidth $W$) over which information on the quasiparticle response can be determined, thus
further constraining theory and allowing the possibility to distinguish between the various phenomenologies proposed.

Historically, the type of approach employed to analyse the optical conductivity data for a particular sample has
depended on where it resides in the cuprate phase diagram. In low-doped cuprates, i.e. near half-filling, it has become
customary to adopt the so-called two-component picture that assumes a Drude component at low frequencies coupled with a
Lorentzian in the mid-infrared region which contains a large fraction of the spectral weight. \cite{BasovTimusk} In
optimally and over-doped cuprates on the other hand, these two components appear to merge, making a one-component model
the more appropriate. In this case, one uses the so-called extended or generalized Drude model \cite{AllenMikkelsen}
that assumes a single Drude component for $\omega < W$  but with a scattering rate $\Gamma$($T$,$\omega$) and coupling
constant $\lambda$($T$,$\omega$) showing strong frequency dependence. In this case,

\begin{eqnarray}
\sigma(T, \omega) = \frac{\Omega_p^2/4\pi}{\Gamma(T, \omega) - i\omega[1 + \lambda(T, \omega)]} \label{eq:one}
\end{eqnarray}

where $\Omega_p$ is the plasma frequency and $\lambda$($T$,$\omega$) is causally related to $\Gamma$($T$,$\omega$) via
the Kramers-Kronig transformation

\begin{eqnarray}
\lambda(T, \omega) = \frac{2}{\pi} P \int_0^{\infty} \frac{\Gamma(T, \Omega)}{\Omega^2 - \omega^2} \textrm{d}\Omega
\label{eq:two}
\end{eqnarray}

Here $P$ stands for the Cauchy principal value. To extract $\Gamma$($T$,$\omega$), it is common practice simply to
invert Eqn. (1), i.e.

\begin{eqnarray}
\Gamma(T,\omega) = \frac{\Omega_p^2}{4\pi} \textrm{Re} \left[ \frac{1}{\sigma(T, \omega)} \right] \label{eq:three}
\end{eqnarray}

\begin{figure}
\includegraphics[width=6.0cm,keepaspectratio=true]{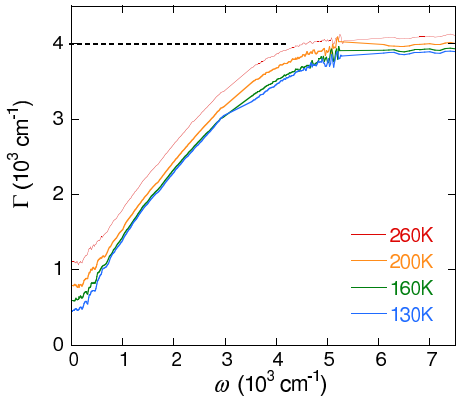}
\caption{(Color online) $\Gamma$($T$,$\omega$) extracted from Eq.(\ref{eq:three}) for optimally doped Y-Bi2212. The
dashed line represents the value $\Gamma_{\rm sat}$ = 4000cm$^{-1}$ $\sim$ 0.5eV at which $\Gamma$($T$,$\omega$)
saturates. Data reproduced from Ref. [11] by kind permission of D. van der Marel.} \label{Figure1}
\end{figure}

Although a consensus has not yet been reached on its overall applicability, \cite{Tanner} the generalized Drude
approach can in principle provide valuable spectroscopic information on $\Gamma$($T$,$\omega$) and
$\lambda$($T$,$\omega$), the optical analogs of the real and imaginary parts of the quasiparticle self-energy. In {\it
optimally doped} cuprates, $\Gamma$($\omega$) extracted in this way is invariably found to be linear in frequency below
3000 cm$^{-1}$. \cite{Basov96, Quijada, Tu, vdM03, MaWang} Such behavior is exemplified by recent state-of-the-art
optical data on optimally doped Bi$_2$Sr$_2$Ca$_{0.92}$Y$_{0.08}$Cu$_2$O$_8$ (Y-Bi2212) reproduced in Figure
\ref{Figure1}. \cite{vdM03} This linear dependence of $\Gamma$($\omega$) mirrors the ubiquitous $T$-linear resistivity
in optimally-doped material that extends in some cases up to 1000K.\cite{GurvitchFiory} Such linearity in both
frequency and temperature is consistent with a marginal Fermi-liquid (MFL) self-energy. \cite{Varma89} It is argued in
Ref. [11] that the Y-Bi2212 data obey quantum critical scaling (though not of the MFL form) and more recently, a
similar but revised scaling analysis has been carried out by the same group for the trilayer compound
Bi$_2$Sr$_2$Ca$_2$Cu$_3$O$_{10}$. \cite{vdM06}

One of the primary objectives of the present work is to sound a note of caution for conclusions drawn using
Eq.~(\ref{eq:three}), the validity of which relies on all parameters in (3) being isotropic. Over the last decade,
evidence has accumulated for a very significant basal-plane anisotropy in cuprates both in the quasiparticle velocity
$v_F$ and its lifetime $\tau$, firstly from measurements of interlayer magnetoresistance \cite{Hussey96} and
subsequently (and more directly) from angle-resolved photoemission spectroscopy (ARPES).\cite{Valla00} Moreover if this
anisotropy is energy dependent, what one actually obtains from plotting $\Gamma$($T$,$\omega$) $\sim$
Re[1/$\sigma$($T$,$\omega$)] is the $T$- and $\omega$-dependent $\Gamma$ embracing a global angular average of the
anisotropic parts of $\Gamma$($\phi$,$T$,$\omega$), $\lambda$($\phi$,$T$,$\omega$), $v_F$($\phi$) and the in-plane
Fermi wave vector $k_F$($\phi$). Hence, all anisotropy {\it together with its frequency dependence} is being subsumed
into $\Gamma$($\omega$). We argue here that a more rigorous way to model the data is to use the fully anisotropic
expression for $\sigma_{ab}$($T$,$\omega$) within an extended Drude formalism. As we shall show, employing the data of
Ref. [11] for illustration, this can have a profound effect upon the extracted frequency dependence of
$\Gamma$($\omega$).

As seen in Fig. \ref{Figure1}, $\Gamma$($\omega$) in Y-Bi2212 starts to deviate from linearity at frequencies above
3000cm$^{-1}$, tending to a constant value $\Gamma_{\rm sat} \sim$ 4000cm$^{-1}$. Such \lq saturation' in
$\Gamma$($T$,$\omega$) is suggestive of strong coupling to bosons rather than critical scaling phenomena, though given
the high frequency at which saturation sets in ($\sim$ 5000 cm$^{-1}$), presumably not of coupling to phonons.
Recently, Norman and Chubukov \cite{NormanChubukov} showed that a model based on coupling to a broad spectrum (there
presumed to be of spin fluctuations) extending out to 0.3eV captures most of the essential features of the data in Ref.
[11], although a gapped MFL model also gave similarly good agreement.

One aspect of the data in Fig. \ref{Figure1} however appears at odds with the standard picture of electron-boson
coupling. According to the standard Allen formalism,\cite{Allen71, Shulga} saturation of $\Gamma$($\omega$) sets in at
progressively higher frequencies as $T$ is raised, more so still if the bosonic response were to broaden and shift to
higher frequencies, as is expected if the strongest coupling is to antiferromagnetic spin fluctuations. The data
however do {\it not} show this tendency; if anything $\Gamma$($\omega$) saturates at a {\it lower} frequency as $T$
increases. This counter trend in $\Gamma$($T$,$\omega$) is seen particularly clearly in the optical response of
underdoped Ca$_{2-x}$Na$_x$CuO$_2$Cl$_2$.\cite{Waku}

We consider an alternative physical origin for the saturation in $\Gamma$($T$,$\omega$), namely an asymptotic approach
to the Mott-Ioffe-Regel (MIR) limit for coherent charge propagation. The MIR criterion states that the electron
mean-free-path $\ell$ has a lower limit of order the interatomic spacing $a$ (alternatively speaking, $\Gamma$ can
never exceed the bare bandwidth $W$).\cite{IoffeRegel, Mott72} Beyond that point, the concept of carrier velocity is
lost and all coherent quasiparticle motion vanishes. Such a threshold is seen, for example, in metals exhibiting
resistivity saturation, where the saturation value is found to be consistent with $\ell$ = $a$.\cite{Gunnarsson,
Hussey04} The MIR limit was in fact invoked to account for the saturation of $\Gamma$($\omega$) in
Ca$_{2-x}$Na$_x$CuO$_2$Cl$_2$.\cite{Waku} In Y-Bi2212, $\Omega_p$ is estimated to be $\sim$ 19500 cm$^{-1}$, which upon
converting to SI units, gives $\langle v_F \rangle \sim$ 3.4 $\times$ 10$^5$ ms$^{-1}$ for $\langle k_F \rangle \sim$
7.0 nm$^{-1}$ (here $\langle v_F \rangle$ and $\langle k_F \rangle$ refer to the angle-averaged bare velocity
$v_F$($\phi$) and Fermi wave vector $k_F$($\phi$) respectively, the latter derived from ARPES measurements
\cite{EschrigNorman, Kaminski05}). Taking the strict definition of the MIR criterion we thus estimate $\Gamma_{\rm
MIR}$ = $\langle v_F \rangle$/$a \sim$ 4500 cm$^{-1}$ (converting back to cgs). Comparison with Figure 1 suggests that
$\Gamma_{\rm sat}$ $\sim$ $\Gamma_{\rm MIR}$, i.e. the saturation value of $\Gamma$($T$,$\omega$) is compatible with
the MIR limit as defined. Although $\Gamma_{\rm sat}$ in Figure 1 does show a small increase with increasing
temperature ($\sim 5\%$ for 100K $\leq T \leq$ 300K), the change, as stated above, is not as much as would be expected
to arise, say, from coupling to bosons.\cite{Hwang06}

The combination of strong four-fold basal plane anisotropy and saturation of the scattering rate at the MIR limit has
previously been incorporated by one of the authors into a phenomenological model, the so-called anisotropic scattering
rate saturation (ASRS) model, to account for a number of anomalous in-plane transport properties of high-$T_c$
cuprates, including the separation of transport and Hall lifetimes (at optimal doping) and modifications to Kohler's
rule.\cite{Hussey03} In this paper, we extend the ASRS model into the frequency domain using the generalized Drude
approach and employ the derived expressions to fit experimental $\sigma_{ab}$($T$,$\omega$) data for optimally doped
Y-Bi2212 over a range of temperatures. The model is found to reproduce the optical response of Y-Bi2212 over two
decades in frequency with parameters that are consistent with those extracted from dc transport measurements on
Bi2212.\cite{Hussey03} Details of the mass enhancement, extracted self-consistently in the analysis, are also compared
with specific heat and ARPES data in Bi-2212 with similar success.

The paper is set out as follows. The ASRS model is introduced in Section II and extended into the realm of finite
frequencies. In Section III, the fitting to the experimental data on Y-Bi2212 is presented whilst in Section IV, the
possible consequences for our understanding of thermodynamic properties in cuprates are discussed. Finally, we offer
our summary and conclusions in Section V.

\section{ASRS model and optical conductivity}
\label{asrs}

The ASRS model assumes a temperature (energy) dependence which is quadratic everywhere on the Fermi surface but
exhibits strong (four-fold symmetric) basal plane anisotropy. Evidence for an (approximately) quadratic temperature
(energy) dependence in cuprates, extending over a wide temperature (energy) range, initially came from measurements of
the inverse Hall angle cot$\theta_{\rm H}$($T$)\cite{Chien91, Carrington92, Manako92} and more recently from ARPES
studies of the self-energy near the nodal region along ($\pi$, $\pi$).\cite{Kordyuk04} When combined with an
appreciable elastic scattering rate that also contains four-fold anisotropy (sometimes ascribed to small angle
scattering off impurities located between the CuO$_2$ planes \cite{AbrahamsVarma}), the intrinsic or \lq ideal'
scattering rate $\Gamma_{\rm ideal}$($\phi$,$T$) can be written most succinctly as

\begin{eqnarray}
\Gamma_{\rm{ideal}}(\phi,T) =  \alpha(1 + c\cos^{2}2\phi) + \beta(1 + e\cos^{2}2\phi)T^{2} \label{eq:four}
\end{eqnarray}

where $c$ and $e$ are anisotropy parameters while $\alpha$ and $\beta$ are coefficients of the component scattering
rates along ($\pi$, $\pi$). To extend the model to finite frequencies we shall simply adopt the standard Fermi-liquid
condition for electron-electron scattering \cite{Gurzhi} and accordingly write $\Gamma_{\rm
ideal}$($\phi$,$T$,$\omega$) as

\begin{eqnarray}
\Gamma_{\rm{ideal}}(\phi,T) = &&\alpha(1 + c\cos^{2}2\phi) +\nonumber\\
&& \beta(1 + e\cos^{2}2\phi) \left(T^{2} + (\hbar\omega/2\pi k_B)^2\right) \label{eq:five}
\end{eqnarray}

To accommodate the MIR limit we invoke the anisotropic \lq parallel-resistor' formula,

\begin{eqnarray}
\frac{1}{\Gamma_{\rm{eff}}(\phi, T, \omega)} = \frac{1}{\Gamma_{\rm{ideal}}(\phi, T, \omega)} +
\frac{1}{\Gamma_{\rm{MIR}}} \label{eq:six}
\end{eqnarray}

to mimic the form of the resistivity $\rho$($T$) found in systems exhibiting resistivity saturation.\cite{Wiesmann} (We
stress here that this formula should be viewed as scattering rates adding in parallel rather than as different
conduction channels). The formula presumes that the MIR limit is manifest at all temperatures and, by extension, all
energies below the bandwidth. To fit to the optical conductivity data, we further need the effective mass enhancement
factor $\lambda_{\rm eff}$($\phi$,$T$,$\omega$), obtained via the appropriate Kramers-Kronig
transformation~(\ref{eq:two}). To simplify our working, we make the following substitutions, $\Gamma_0$ = $\alpha$(1 +
$c$cos$^2$2$\phi$), $\Theta$ = $\beta$(1 + $e$cos$^2$2$\phi$), $\hbar$/2$\pi k_B$ = 1 and $\Lambda$ = $\Gamma_{\rm
MIR}$ + $\Gamma_0$ + $\Theta T^2$, and then re-arrange $\Gamma_{\rm eff}$ as

\begin{eqnarray}
\Gamma_{\rm{eff}}(\phi, T, \omega) = \Gamma_{\rm{MIR}} - \left(\frac{\Gamma_{\rm{MIR}}^2}{\Lambda + \Theta\omega^2}
 \right) \label{eq:seven}
\end{eqnarray}

Thus $\lambda_{\rm eff}$($\phi$,$T$,$\omega$) becomes:

\begin{eqnarray}
&&\lambda_{\rm{eff}}(\phi, T, \omega) = \frac{2}{\pi} P \int_0^{\infty} \frac{\Gamma_{\rm{MIR}}}{\Omega^2 - \omega^2}
d\Omega \nonumber\\
&& - \frac{\Gamma_{\rm{MIR}}^2}{\Lambda} \frac{2}{\pi} P \int_0^{\infty} \frac{1}{(\Omega^2 - \omega^2)}\frac{1}{(1 +
\Theta\omega^2/\Lambda)} d\Omega \label{eq:eight}
\end{eqnarray}

The first integral here is zero. The second gives

\begin{eqnarray}
\lambda_{\rm{eff}}(\phi, T, \omega) = \Gamma_{\rm{MIR}}^2 \left(\frac{\Theta}{\Lambda}\right)^{1/2} \left(
\frac{1}{\Lambda + \Theta\omega^2} \right) \label{eq:nine}
\end{eqnarray}

A similar ansatz (but without anisotropy) has been used previously to replicate the form of $\Gamma$($\omega$)
extracted from an extended Drude analysis for heavy fermion compounds \cite{Sulewski} although in that case, no
physical explanation for $\Gamma_{\rm MIR}$ was provided. Interestingly, when we extrapolate to $\omega$ = 0 and low
$T$ where $\Lambda \sim \Gamma_{\rm  MIR}$, Eq.~(\ref{eq:seven}) reduces to an expression for the dc mass enhancement
factor, $\lambda_{\rm eff}$(0) $\sim$ ($\hbar \Theta \Gamma_{\rm MIR}$/4$\pi^2k_B$)$^{1/2}$ $\sim$ ($\hbar \Theta
\langle v_F \rangle$/4$\pi^2ak_B$)$^{1/2}$. This expression (in its isotropic form) has been shown to give a reliable
estimate of the $A$ coefficient of the $T^2$ resistivity found in a variety of strongly correlated
metals.\cite{Hussey05}

\begin{figure}
\includegraphics[width=6.0cm,keepaspectratio=true]{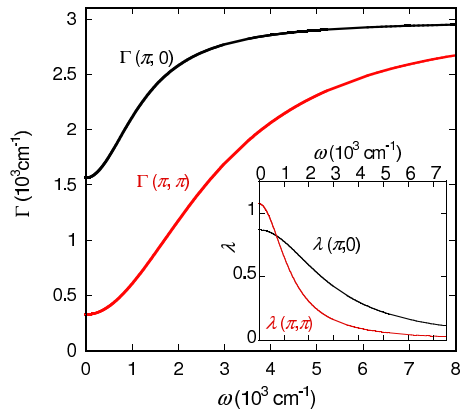}
\caption{(Color online) $\Gamma_{\rm eff}$($\omega$) along ($\pi$, $\pi$) (red) and ($\pi$, 0) (black) at $T$ = 200K
according to the ASRS phenomenology. Inset: Corresponding $\lambda_{\rm eff}$($\omega$) for the same two orientations,
obtained via the appropriate Kramers-Kronig transformation.} \label{Figure2}
\end{figure}

Figure \ref{Figure2} shows the evaluated frequency-dependent ASRS scattering rate $\Gamma_{\rm eff}$($\omega$) along
($\pi$, $\pi$) (red line) and ($\pi$, 0) (black line). (In this example, the parameters used are the same as those used
in Figure \ref{Figure3} below to fit the 200K data). The corresponding mass enhancement factors $\lambda_{\rm
eff}$($\omega$) are shown in the inset. One should note that even though the anisotropy in $\Gamma_{\rm ideal}$($\phi$)
is large ($e$ = 9 in the example shown), the effective anisotropy in $\Gamma_{\rm eff}$($\omega$=0) is significantly
less so ($\sim$ 5). Moreover the anisotropy in $\lambda_{\rm eff}$($\omega$=0) is inverted with respect to $\Gamma_{\rm
eff}$($\omega$=0). Both of these features are due to the presence of $\Gamma_{\rm MIR}$ which acts to \lq dampen' the
intrinsic anisotropy in the interaction strength. Note too that $\Gamma_{\rm eff}$($\omega$) along ($\pi$, 0) saturates
at a much lower frequency than occurs along ($\pi$, $\pi$) and at high frequencies, the two scattering rates converge,
as one should expect through the averaging of $k$-states.

Finally, we insert (5), (6) and (9) into the full expression for the real-part of the conductivity
$\sigma_1$($T$,$\omega$) = Re[$\sigma_{ab}$($T$,$\omega$)] for a two-dimensional Fermi surface where

\begin{eqnarray}
\sigma_{ab}(T, \omega) = &&\frac{e^2}{4\pi^3\hbar} \left(\frac{2\pi}{d} \right)\int_0^{2\pi} \frac{k_F(\phi) v_F(\phi)
\rm{cos}^2(\phi - \gamma)}{\rm{cos}\gamma} \nonumber\\
&& \times \frac{d\phi}{\Gamma_{\rm{eff}}(\phi, T, \omega) - i\omega[1 + \lambda_{\rm{eff}}(\phi, T, \omega)]}
\label{eq:ten}
\end{eqnarray}

(For a derivation of this expression in the $\omega$ = 0 limit see the Appendix of Ref. [29]). Here $d$ is the lattice
spacing and $\phi$ is taken around the two-dimensional projection of the CuO$_2$ Fermi surface. $\gamma$ =
tan$^{-1}$[$\partial$/$\partial\phi$(log$k_F$($\phi$))] is the angle between $v_F$ and d$k_r$ (the infinitesimal vector
element along $k_F$). Whilst this Boltzmann-type approach cannot be applicable over the entire energy range, we believe
the formalism presented here adequately represents the main issue regarding anisotropy and that the results will not
differ significantly from those which would be extracted via a more precise Kubo formalism.\cite{Allen04}

\section{Results}
\label{results}

\begin{figure}
\includegraphics[width=6.0cm,keepaspectratio=true]{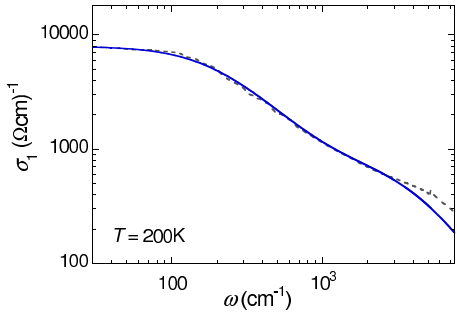}
\caption{(Color online) $\sigma_1$($\omega$) data (dashed grey line) and fit (solid blue line) at $T$ = 200K. Fitting
parameters are $\alpha$ = 64 cm$^{-1}$, $\beta$ = 0.0072 cm$^{-1}$/K$^2$, $c$ = 3 and $e$ = 9.} \label{Figure3}
\end{figure}

In order to minimise the number of fitting parameters, we have fixed both $k_F$($\phi$) and $v_F$($\phi$) using the
tight-binding expression\cite{EschrigNorman} for the CuO$_2$ plane (bonding) band in Bi2212\cite{Note} and set
$\Gamma_{\rm MIR}$ = $\langle v_F \rangle$/$a$. Hence the only free parameters in our fitting procedure are the
coefficients in our expression for $\Gamma_{\rm ideal}$($\phi$, $T$, $\omega$), namely $\alpha$ and $\beta$, $c$ and
$e$. Figure \ref{Figure3} shows our fit to $\sigma_1$($\omega$) at $T$ = 200K. The fitting parameters are listed in the
Figure caption. The ASRS parameterization provides an excellent fit to the experimental data from the lowest
frequencies studied up to $\omega \sim$ 3000 cm$^{-1}$. Note that no scaling factors were applied in this fitting
procedure. The deviation of the fit above 3000 cm$^{-1}$ is most probably due to the fact that the magnitude of
$\langle v_F \rangle$ from tight-binding which was used to fix $\Gamma_{\rm MIR}$ is somewhat lower than that estimated
from the optical data.

\begin{figure}
\includegraphics[width=6.0cm,keepaspectratio=true]{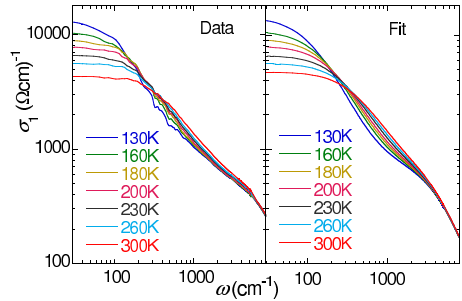}
\caption{(Color online) $\sigma_1$($\omega$) data (left panel) and fit (right panel) for a range of temperatures 130K
$\leq T \leq$ 300K. In this fitting routine, all fitting parameters were fixed to their value(s) at 200K.}
\label{Figure4}
\end{figure}

To track the evolution of $\sigma_1$($T$,$\omega$) for all $T > T_c$, we simply insert the relevant temperature
into~(\ref{eq:five}) keeping all other parameters in~(\ref{eq:ten}) fixed. The fittings are shown in Figure
\ref{Figure4}. The evolution of the low-frequency response in particular is reproduced well by the parameterization.
Similarly good fits were obtained too for $\sigma_2$($T$,$\omega$), the imaginary part of the conductivity and also for
the phase-angle $\varphi$ (not shown), though given their mutual inter-dependence via the Kramers-Kronig relations,
this is not so surprising.

The dc transport properties can be derived within the same parameterization scheme,\cite{Hussey03}

\begin{eqnarray}
\rho_{ab}(T) = \frac{1}{\sigma_{ab}(T, \omega = 0)} \label{eq:eleven}
\end{eqnarray}

\begin{eqnarray}
\tan\Theta_H (T) = \frac{\sigma_{xy}(T)}{\sigma_{ab}(T, \omega = 0)} \label{eq:twelve}
\end{eqnarray}

where
\begin{eqnarray}
\sigma_{xy}(T) = && \frac{-e^3B}{4\pi^3\hbar^2} \left(\frac{2\pi}{d} \right)\int_0^{2\pi} \frac{v_F(\phi) \cos(\phi -
\gamma)}{\Gamma_{\rm eff}(\phi, T,\omega=0)} \nonumber\\
&& \times \frac{\partial}{\partial\phi} \left(\frac{v_F(\phi) \sin(\phi - \gamma)}{\Gamma_{\rm eff}(\phi,
T,\omega=0)}\right)\textrm{d}\phi \label{eq:thirteen}
\end{eqnarray}

\begin{figure}
\includegraphics[width=6.0cm,keepaspectratio=true]{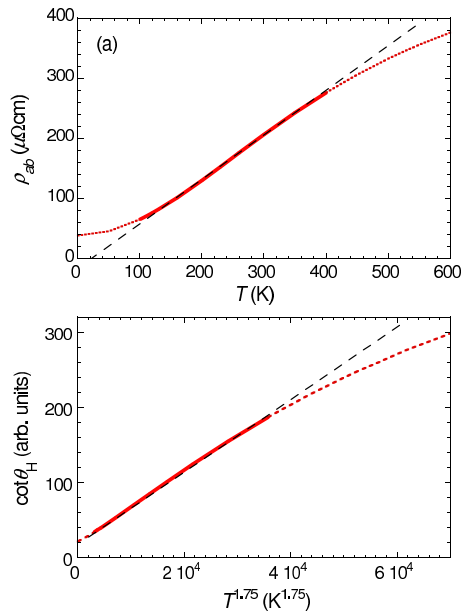}
\caption{(Color online) (a) $\rho_{ab}$($T$) extracted from the fitting parameters listed in the Figure caption of Fig.
\ref{Figure3}. (b) Corresponding cot$\theta_{\rm H}$($T$) versus $T^{1.75}$ for the same fitting parameters. The thick
solid lines highlight the region between 100K and 400K where $\rho_{ab}$($T$) is $T$-linear and cot$\theta_{\rm
H}$($T$) $\sim$ $A$ + $BT^{1.75}$. The dashed lines are guides to the eye.} \label{Figure5}
\end{figure}

Figure \ref{Figure5} shows the resultant simulation plots of $\rho_{ab}$($T$) and cot$\theta_{\rm H}$($T$) in the
temperature interval 0 $\leq T \leq$ 600K.  The region between 100K and 400K is shown as a thick solid line for
emphasis. Over this latter temperature range $\rho_{ab}$($T$) is found to vary linearly with temperature (as indicated
by the dashed line) but displays clear upward curvature at lower temperatures as $\rho_{ab}$($T$) approaches a finite
{\it positive} intercept. Above 400K, $\rho_{ab}$($T$) starts to deviate from linearity as it approaches the saturation
value assumed by the model. The fact that $\rho_{ab}$($T$) in optimally doped cuprates follows a $T$-linear dependence
up to much higher temperatures and values {\it beyond} the MIR limit, without showing any sign of saturation, may
appear at first sight to invalidate the model. Optical conductivity measurements in cuprates at elevated temperatures
however have shown evidence for a shift in the low-frequency spectral (Drude) weight above 300K suggesting some form of
thermally-induced decoherence of the quasiparticles.\cite{Merino00, Hussey04} The vanishing of the zero-frequency Drude
peak also implies that the continuation of the $T$-linear dependence of $\rho_{ab}$($T$) beyond the MIR limit cannot be
associated with an unbounded escalation of the ($T$-linear) scattering rate. This conjecture is further supported by
the observation of saturation of $\Gamma$($T$,$\omega$) at high frequencies at a value $\Gamma_{\rm sat}$ $\sim$
$\Gamma_{\rm MIR}$ $\sim W$ (see Fig. \ref{Figure1}).

In contrast to the $T$-linearity of $\rho_{ab}$($T$), cot$\theta_{\rm H}$($T$) is found to follow the form $A$ + $BT^n$
between $T_c$ and 400K, with $n$ = 1.75. This unusual power law is remarkably similar to that observed experimentally
($n$ = 1.78) in thin films of optimally doped Bi2212.\cite{Konstantinovic} The origin of the so-called \lq separation
of lifetimes' \cite{Anderson91} governing $\rho_{ab}$($T$) and cot$\theta_{\rm H}$($T$) in cuprates has been a
long-standing controversy.\cite{HusseyReview} In single lifetime models with a strongly anisotropic $\ell$({\bf k}),
the Hall conductivity as given in Eq.~(\ref{eq:thirteen}) is dominated by those quasiparticles with the longest
mean-free-path.\cite{Ong91} In optimally doped cuprates, the $T$-dependence of cot$\theta_{\rm H}$ would thus be
determined by the nodal quasiparticles near ($\pi$, $\pi$) which present a near-quadratic temperature (and frequency)
dependence. $\rho_{ab}$($T$) on the other hand, is a global average of $\Gamma_{\rm eff}$($\phi$,$T$) and since the
anti-nodal saddle regions near ($\pi$, 0) exhibit a {\it sub}-linear, almost flat $T$-dependence, the integral over the
entire Fermi surface would yield $\rho_{ab}$($T$) $\sim T$. Comparable considerations led Ioffe and Millis to develop
their widely regarded \lq cold-spots' model for high-$T_c$ cuprates. \cite{IoffeMillis}

\begin{figure}
\includegraphics[width=6.0cm,keepaspectratio=true]{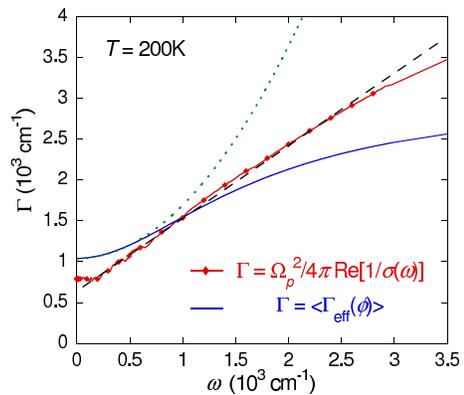}
\caption{(Color online) Comparison between $\Gamma$($\omega$) obtained from Eq.~(\ref{eq:three}) (solid red line with
diamonds) and the angle-averaged $\Gamma_{\rm eff}$($\omega$) = (1/2$\pi$) $\oint \Gamma_{\rm
eff}$($\phi$,$\omega$)d$\phi$ extracted from the fitting parameters (solid blue line). The dotted and dashed lines are
quadratic and linear extrapolations up from the respective low-frequency limits.} \label{Figure6}
\end{figure}

Let us now examine how the anisotropy becomes reflected in the form of $\Gamma$($\omega$) and vice versa. In Figure
\ref{Figure6}, the solid red line with diamonds represents $\Gamma$($\omega$) at $T$ = 200K as determined by
Eq.~(\ref{eq:three}), i.e. just the inversion of the (isotropic) conductivity, below 3500 cm$^{-1}$. As described in
Ref. [11] {\it and many other papers adopting the same procedure}, $\Gamma$($\omega$) appears linear below 3000
cm$^{-1}$ (thick dashed line). The solid blue line in Figure \ref{Figure6} is the angle-averaged $\langle \Gamma_{\rm
eff}$($\omega$)$\rangle$ = (1/2$\pi$) $\oint \Gamma_{\rm eff}$($\phi$,$\omega$)d$\phi$ where $\Gamma_{\rm
eff}$($\phi$,$\omega$) is the ASRS scattering rate inserted into the full anisotropic expression~(\ref{eq:ten}) along
with $\lambda_{\rm eff}$($\phi$,$\omega$) and the anisotropic Fermi surface parameters. Note now that $\Gamma_{\rm
eff}$($\omega$) $\propto \omega^2$ up to 800 cm$^{-1}$ (dotted line) and that the $\omega$-linear regime is restricted
to a narrow crossover region between the $\omega^2$ low-frequency response and the onset of saturation. The key point
to make here is that {\it both} parameterizations, the one isotropic, the other anisotropic, can fit the optical
conductivity data equally well in this frequency range, and yet the form of $\Gamma$($\omega$) is markedly different in
the two cases. Of course, our modelling does not rule out a contribution to $\Gamma$($\omega$) that is intrinsically
$\omega$-linear; it merely highlights the fact that when neglecting anisotropy one is going to infer information about
$\Gamma$($\omega$) solely from the ($\omega$-linear) frequency dependence of Re[1/$\sigma$($\omega$)], without any
physical justification for doing so.

\section{Discussion}
\label{discussion}

\begin{figure}
\includegraphics[width=6.0cm,keepaspectratio=true]{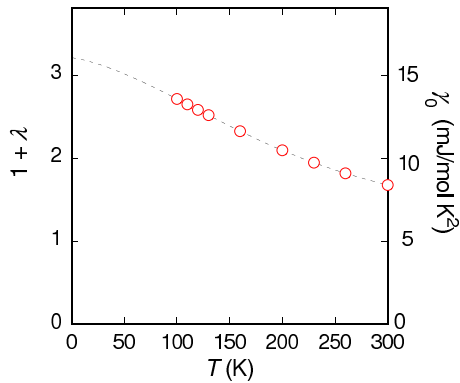}
\caption{(Color online) $T$-dependence of the mass enhancement 1 + $\lambda_{\rm eff}$ as extracted from the fitting
parameters of Model 1. The axis labels on the right hand side correspond to the electronic specific heat coefficient
$\gamma_0$ assuming a band mass $m_b$ = 1.} \label{Figure7}
\end{figure}

In this section we discuss some of the implications of the above modelling beyond the inherent form of
$\Gamma$($\omega$). One immediate consequence of our assumption that $\Gamma_{\rm MIR}$ constitutes a ceiling on
$\Gamma$($\omega$) is the reduction in $\lambda_{\rm eff}$(0) with increasing temperature. Due to the Kramers-Kronig
relation, $\lambda_{\rm eff}$(0) is governed to a large degree by the {\it difference} between the low and high
frequency limits of $\Gamma$($\omega$). As the temperature rises, $\Gamma$($\omega$=0) quickly rises due to the
increase in inelastic scattering. $\Gamma_{\rm MIR}$ on the other hand is independent of temperature (if one ignores
the effects of thermal expansion and $T$-induced variations in the band dispersion). $\Gamma$($\omega$) is thus
confined by these two limiting extremes, leading to an overall reduction in $\lambda_{\rm eff}$(0). Since $\lambda_{\rm
eff}$(0) governs the mass-enhancement, the effect should be manifest in thermodynamic properties such as the electronic
specific heat coefficient $\gamma_0$($T$). Figure \ref{Figure7} shows the $T$-dependence of the (angle-averaged)
$\lambda_{\rm eff}$(0) and $\gamma_0$ derived from our modelling of the optical data assuming a band mass $m_b$ = 1.
$\gamma_0$($T$) is seen to rise by approximately 50\% as the temperature falls from 300K to 100K. Both the magnitude of
$\gamma_0$ and its variation with temperature are in reasonable agreement with measured data on Bi2212.\cite{Loram}
Within the standard Allen formalism,\cite{Shulga} $\lambda_{\rm eff}$(0) should be essentially $T$-independent if
phonons provide the dominant interaction (otherwise $\rho$($T$) in a standard metal would deviate from its strictly
$T$-linear temperature dependence). Within a spin fluctuation picture, one could expect by contrast $\lambda_{\rm
eff}$(0) and hence $\gamma_0$($T$) to fall with increasing temperature as the low-frequency susceptibility is
wiped-out. However, as we have already argued, within the same scenario one should expect the onset of saturation in
$\Gamma$($\omega$) to shift to higher frequencies with increasing temperature, which to our knowledge has never been
observed experimentally.

One can compare also the mass enhancement of the optical (particle-particle) self-energy to that of the single-particle
self-energy $\Sigma$($\omega$) derivable from ARPES along the different $k$-space directions. Along ($\pi$, $\pi$), for
example, $\partial$Re$\Sigma$/$\partial\omega \sim$ 1.9 for slightly underdoped Bi-2212 at $T$= 130K.\cite{Kordyuk05}
According to our fitting of the optical data, we obtain $\lambda_{\rm eff}$(0) = 1.5 along ($\pi$, $\pi$) at the same
temperature, giving 1 + $\lambda_{\rm eff}$(0) $>$ $\partial$Re$\Sigma$/$\partial\omega$ at ($\pi$, $\pi$). This
inequality suggests that (current-current) vertex corrections do play some role in the physics of the nodal regions.
What the ARPES data of Ref. [45] do {\it not} show is the tendency towards saturation in Im$\Sigma$($\omega$), a key
component of the ASRS phenomenology. However, a recent ARPES study of the single-layer Bi-compound
Bi$_{1.74}$Pb$_{0.38}$Sr$_{1.88}$CuO$_{6+\delta}$ has revealed that at higher energies, there is evidence for
saturation of Im$\Sigma$($\omega$) at values comparable to the bandwidth.\cite{Xie} It would be interesting in due
course to examine whether the phenomenology presented here could also be applied to explain the form of the
single-particle self-energy.

\section{Conclusions}
\label{conclusions}

In this paper, we have sought to counter the ubiquitous use of Eq.~(\ref{eq:three}) in extraction of $\Gamma$($\omega$)
from the in-plane normal-state optical data of high-$T_c$ cuprates. We have also demonstrated how basal-plane
anisotropy in the scattering rate, ignored until now in the analysis of $\sigma_{ab}$($\omega$), can bring about a
marked change in the analytical form of $\Gamma$($\omega$); most strikingly, the customary linear frequency dependence
at low $\omega$ is seen as being an artefact of a fitting routine that presumes isotropic scattering. The key take-home
message is that the intrinsic low energy response of optimally doped cuprates is only able to be properly determined
once all anisotropic factors are fully taken into account. Because the system contains very significant anisotropy
which may vary with frequency, one cannot use~(\ref{eq:three}) ubiquitously to extract $\Gamma$($\omega$) from
$\sigma_{ab}$($\omega$). We have argued that a more rigorous way to proceed is to input a form of
$\Gamma$($\phi$,$T$,$\omega$) into the full anisotropic expression for $\sigma_{ab}$($\omega$) which in conjunction
with its Kramers-Kronig partner $\lambda$($\phi$,$T$,$\omega$) and the correct Fermi surface parameterization would fit
the data.

Above we have proposed one such parameterization, based on a $\Gamma$($\omega$) with a quadratic frequency dependence,
strong basal-plane anisotropy and a tendency towards saturation at the MIR limit, which secures a good account of the
optical response. Moreover, we have indicated how the same parameterization can account for other physical properties
such as anomalies in the dc transport and the specific heat. The ASRS phenomenology presented here clearly is not the
whole story however. Whilst there is a growing body of evidence for saturation and a quadratic frequency dependence in
cuprates, particularly along the nodal direction, there is mounting evidence too for some form of strong bosonic
feature in the anti-nodal regions near the Brillouin zone boundaries,\cite{Kim03, Cuk, AbdelJawad} the origin of which
remains controversial.\cite{Wilson04, Deveraux04, Borisenko03} This contribution to $\Gamma$($\phi$,$T$,$\omega$)
should also be explored within the present phenomenology for completeness. What we would argue however is that since
scattering is already so intense at ($\pi$, 0), the true, \lq ideal' form of $\Gamma$($\omega$) (and
Im$\Sigma$($\omega$)) in this region of the Brillouin zone inevitably will be significantly renormalized due to the
overarching presence of $\Gamma_{\rm MIR}$, masking the inherent nature of $\Gamma$($\omega$) in many of the physical
properties that are measured. This may help to explain why it has taken the community so long to reach a consensus on
the various interactions and scattering mechanisms that influence the self-energy of the in-plane anti-nodal
quasiparticles in the cuprates and which ultimately, may drive high temperature superconductivity. Finally, for a
complete formulation of the physics of cuprates to emerge, a number of the issues raised in the present work will have
to be addressed. In particular, we have shown that saturation of the frequency dependent scattering rate at the MIR
limit greatly influences the transport behavior in cuprates over a very wide energy scale and to acknowledge its
presence may turn out to be a key step in the development of a coherent description of the charge dynamics there.

\begin{acknowledgments}
The authors would like to acknowledge stimulating and enlightening discussions with N. Bontemps, A. V. Chubukov, R. P.
S. M. Lobo, D. B. Tanner, T. Timusk and D. van der Marel. The authors would especially like to thank D. van der Marel
for making his optical data available to us for this analysis and J. A. Wilson for a critical reading of the
manuscript.
\end{acknowledgments}


\begin{thebibliography}{32}
\bibitem{Note} The tight-binding expressions can be approximated by $k_F$($\phi$) = 7.0(1 - 0.05cos4$\phi$ +
0.007cos8$\phi$) nm$^{-1}$ and $v_F$($\phi$) = 2.25(1 - 0.63cos4$\phi$ - 0.05cos8$\phi$) x 10$^5$ ms$^{-1}$. Note the
corresponding $\langle v_F \rangle$ is lower than that derived from the optical plasma frequency by a factor of 0.75.
Tight-binding fits also tend to overestimate the true anisotropy in $v_F$($\phi$) and so for completeness, we also
considered a second parameterization using $v_F$($\phi$) = 2.5(1 + 0.5sin$^2$2$\phi$) $\times$ 10$^5$ ms$^{-1}$ that is
closer to the real $v_F$($\phi$) inferred from ARPES (see, for example, Ref. [26]). The parameters obtained from the
two models howerver turn out to be very similar implying that none of the conclusions are significantly affected by our
choice of $v_F$($\phi$).
\end{thebibliography}


\end{document}